\documentstyle[aps,prl,graphicx,multicol,epsfig,rotate]{revtex}

\begin{document}

\draft \title{Optical Trapping and Manipulation of Nano-objects
with an Apertureless Probe}

\author{Patrick C. Chaumet$^1$, Adel Rahmani$^2$, and Manuel
Nieto-Vesperinas$^3$}

\address{$^1$Institut Fresnel (UMR 6133), Facult\'e des Sciences et
Techniques de St J\'er\^ome, F-13397 Marseille cedex 20, France\\
$^2$Atomic Physics Division, National Institute of Standards and
Technology, Gaithersburg, Maryland 20899-8423\\ $^3$Instituto de
Ciencia de Materiales de Madrid, CSIC, Campus de Cantoblanco Madrid
28049, Spain}

\maketitle

\begin{abstract}

We propose a novel way to trap and manipulate nano-objects above a
dielectric substrate using an apertureless near-field probe. A
combination of evanescent illumination and light scattering at the
probe apex is used to shape the optical field into a localized, three
dimensional optical trap.  We use the coupled-dipole method and the
Maxwell stress tensor to provide a self-consistent description of the
optical force, including retardation and the influence of the
substrate. We show that small objects can be selectively captured and
manipulated under realistic conditions.

\end{abstract}


\begin{multicols}{2}

Since the seminal work of Ashkin on radiation pressure, the
possibility to exploit the mechanical action of optical fields to trap
and manipulate neutral particles has spawned a wide range of
applications.  From atomic and nonlinear physics to biology, optical
forces have provided a convenient way to control the dynamics of small
particles (see Ref.[\ref{ashkin97}] for a review).  Optical tweezers,
for example, have proved useful not only for trapping particles, but
also for assembling objects ranging from microspheres to biological
cells \cite{assembling}. However, most of these manipulations involve
particles whose size is between one and several micrometers. While for
much smaller particles, such as atoms or molecules, the scanning
tunneling microscope provides a powerful tool for manipulation and
engineering \cite{stm}, dealing with neutral particles of a few
nanometers requires new experimental approaches.

The idea of using a metallic probe to trap small particles was
reported by Novotny et al \cite{novotny}. Their calculations showed
that strong field enhancement from light scattering at a gold tip
could generate a trapping potential deep enough to overcome Brownian
motion and capture a nanometric particle in water (a related work by
Okamoto and Kawata demonstrates theoretically the trapping of a
nanometric sphere in water near an aperture probe ~\cite{okamoto}).
This technique should be delicate to implement in practice for three
reasons. First, before the particle can be captured, Brownian
fluctuations will have a disruptive effect. Second, radiation pressure
from the illuminating laser will impart momentum to the
particle. Therefore, one would have to capture a moving
object. Finally, it will be rather difficult to use a near-field probe
to find in water a particle a few nanometers in size. One might wait
for a particle to wander in the trapping region, but such an operating
mode does not allow for selective capture.
 
In this letter, we propose an experimental scheme to selectively
capture and manipulate nanoparticles in vacuum or air above a
substrate, using the tungsten probe of an apertureless near-field
microscope. The particles are not in a liquid environment, hence there
is no Brownian motion and the apertureless probe can be used as a
near-field optical probe to localize and select the
particles~\cite{zenhausern,defornel}.  This is an important asset when
different particles have to be placed according to a specific pattern
or when interactions between particles are investigated.  For example,
with our configuration quantum-dot nanocrystals (CdSe/ZnS nanospheres
a few nanometers in diameter) could be placed in a specific geometry
to study dot-dot interactions, or one nanocrystal could be isolated to
study single-dot properties~\cite{michler00}.

\begin{figure}[H]  
\begin{center}
\includegraphics*[draft=false,width=80mm]{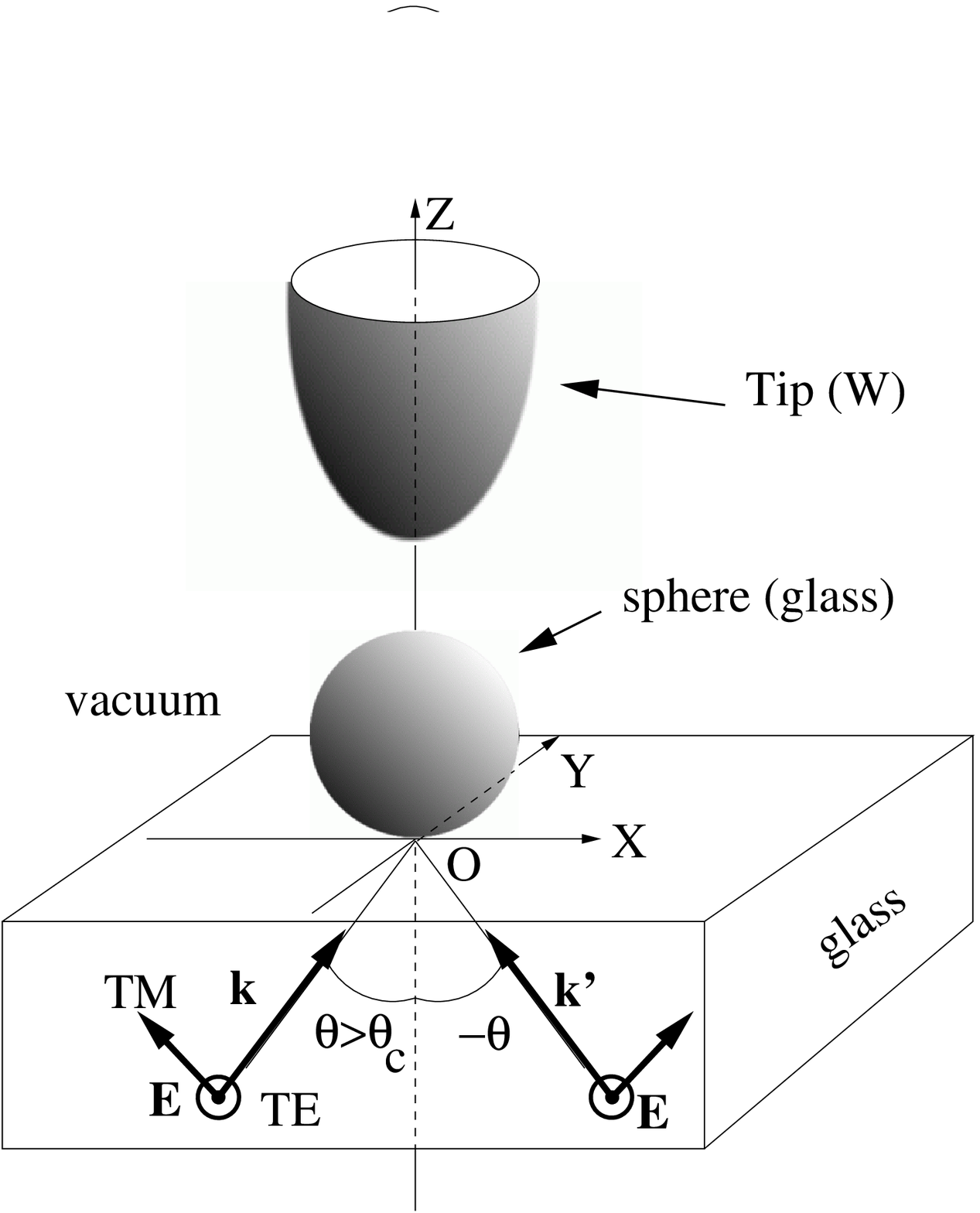}
\end{center}
\caption{Scheme of the configuration. A dielectric sphere (radius
10~nm) on a flat dielectric surface is illuminated under total
internal reflection. A tungsten probe is used to create an optical
trap.}
\end{figure}

We first consider a spherical particle with radius 10~nm placed in air
on a substrate (Fig.~1). Unless it is stated otherwise the
permittivity of both the particle and the substrate is
$\epsilon=2.25$.  The particle is illuminated (wavelength 500~nm) by
two counter-propagating evanescent waves created by the total internal
reflection of plane waves at the substrate/air interface (angle
$\theta >\theta_c=41.8^{\circ}$ with
$\sqrt{\epsilon}\sin\theta_c=1$). These two waves have the same
polarization and a random phase relation.\cite{phase} This symmetric
illumination ensures the lateral force vanishes when the sphere is
just below the tip, thus avoiding that the sphere be pushed away from
the tip.

We study the interactions between the particle and a tungsten probe
(commonly used in apertureless microscopy) with a radius at the apex
of 10~nm. The theory used to compute the optical forces has been
presented in detail elsewhere \cite{chaumet2}. Here we give a succinct
summary. We use the coupled-dipole method~\cite{purcell,chaumet1} to
model the light scattering and find the electromagnetic field inside
the tip and the particle. Note that this procedure takes into account
the multiple scattering between the sphere, the tungsten tip, and the
substrate. We then use the Maxwell stress tensor technique to derive
the optical forces~\cite{stratton}. We emphasize that the stress
tensor approach is exact and does not rely on any assumption regarding
the nature of the field (whether evanescent or propagating) or of the
objects.

\begin{figure}[H]  
\begin{center}
\includegraphics*[draft=false,height=80mm,angle=90]{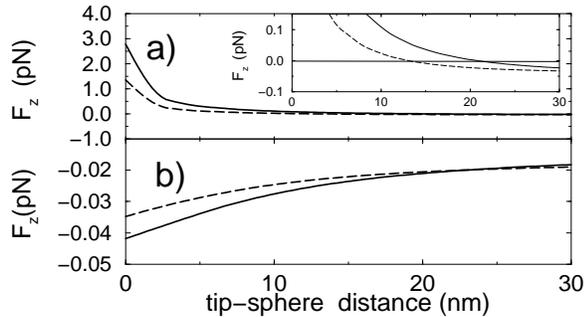}
\end{center}
\caption{$z$ component of the force experienced by the sphere versus
the distance between the tip and the sphere. Solid line :
$\theta=43^{\circ}$. Dashed line : $\theta=50^{\circ}$. a) TM
polarization. b) TE polarization.  }
\end{figure}

All forces are computed for an irradiance of 0.05W/$\mu$m$^2$
(this corresponds, for example, to a 5W laser beam focused over 
an area of 100$\mu$m$^2$). Figure 2 shows
the $z$ component of the force experienced by the sphere versus the
vertical position of the tip above the sphere, for both TE and TM
polarizations and for two angles of illumination ($\theta=43^{\circ}$
and $\theta=50^{\circ}$).  As the tip moves closer to the sphere, the
evolution of the force for the two polarizations is radically
different. For TM illumination there is a large enhancement of the
field near the apex of the probe because of the discontinuity of the
$z$ component of the electric field across the air/tungsten
boundary~\cite{novotny}. This enhancement is responsible for the force
being positive at short distances.  Note that for the sphere, the
force of gravity is on the order of $10^{-7}$pN.  The $z$ component of
the force experienced by the sphere (when the tip is in contact with
the sphere) is about $3$pN which is $10^8$ times its
weight. Hence, gravity can be neglected. Fig. 2a shows that for TM
polarization the force is larger for $\theta=43^{\circ}$ than for
$\theta=50^{\circ}$. This is related to the slower decay of the
evanescent wave for the smaller angle, which results in a weaker
coupling between the sphere and the substrate.  Moreover, the slower
the decay of the evanescent field, the larger the field that reaches
the tip, and the larger the (positive) gradient force caused by the
field enhancement at the apex of the probe. As a consequence, when the
tip approaches the sphere, the sign reversal (negative to positive) of
the $z$ component of the force occurs at a larger distance for
$\theta=43^{\circ}$ ($z$=21~nm) than for $\theta=50^{\circ}$
($z$=13~nm). This is shown in the inset of Fig. 2a. On the other hand,
for TE polarization (Fig.~2b), the magnitude of the $z$ component of
the force increases while the force remains negative (directed toward
the substrate and away from the tip) as the tip gets closer to the
particle.  This prevents any trapping.  Our calculations show that
this behavior is caused by the decrease of the field inside the upper
part of the sphere which, in turn, causes a decrease of the gradient
force.  Another way of explaining this is to note that because the
apex of the tip and the sphere are small compared to the wavelength,
they can be approximated by two dipoles. For TE polarization these
dipoles are essentially parallel to the substrate and as shown in
Ref.[\ref{chaumet3}], two parallel dipoles tend to repel each other.
Since the magnitude of the dipole is stronger for $\theta=43^{\circ}$
(due to the larger intensity of the field), the repulsive force is
also stronger. Note that for a silver tip at the plasmon resonance
frequency, the force acting on the sphere is positive for both
polarizations. At the frequency considered here, tungsten behaves like
an absorbing dielectric and the force is positive only for TM
polarization.

\begin{figure}[H]  
\begin{center}
\includegraphics*[draft=false,width=80mm]{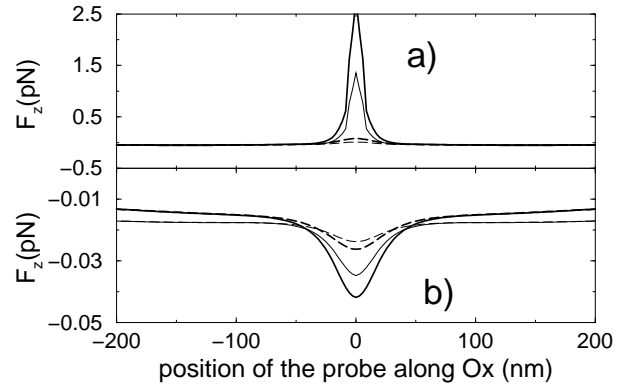}
\end{center}
\caption{Force along $z$ experienced by the sphere as a function of
the lateral position of the probe. The sphere is at the origin. a) TM
polarization, b) TE polarization.  Thick lines :
$\theta=43^{\circ}$. Thin lines : $\theta=50^{\circ}$.  The tip is
either 20~nm (solid lines) or 31~nm (dashed lines) above the
substrate.}
\end{figure}
\begin{figure}[H]  
\begin{center}
\includegraphics*[draft=false,width=80mm]{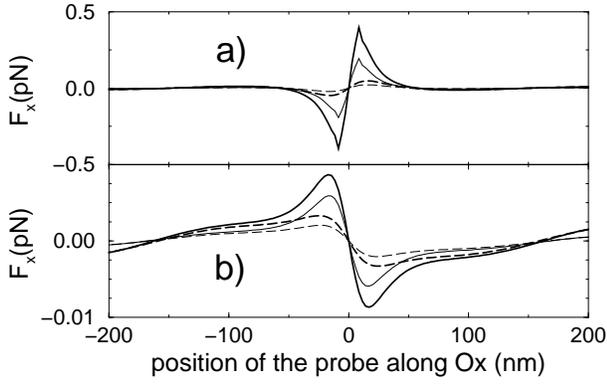}
\end{center}
\caption{Same as Fig.~3 for the force along $x$.}
\end{figure}

To fully assess the probe-particle coupling we study the evolution of
the force experienced by the particle as the probe is moved laterally,
the sphere remaining fixed. The components of the force acting on the
sphere are plotted in Figs. 3 and 4 for two distances between the tip
and the substrate (20~nm and 31~nm), and for two angles of
illumination ($43^{\circ}$ and $50^{\circ}$).  For TM polarization
(Fig. 3a), the $z$ component of the force is negative when the tip is
far from the particle. When the tip gets closer, the particle starts
experiencing a positive force along $z$. The change of sign occurs
between $|x|=$30~nm for $\theta=43^{\circ}$ and $z=20$~nm, and
$|x|=$7~nm for $\theta=50^{\circ}$ and $z=$31~nm.  For TE polarization
the force is negative (Fig. 3b).  Similarly, we plot in Fig. 4 the
lateral force. The symmetry of the force plot is a consequence of the
symmetric illumination.  For TM polarization (Fig. 4a) this force
tends to push the particle toward the tip. For example, if the tip is
located at $x=$10~nm, we can see in Fig. 4a that the $x$ component of
the force experienced by the sphere is positive, hence the lateral
force pushes the sphere toward the tip. If the tip is located at
$x=$-10~nm, the $x$ component of the force is negative and, again, it
pushes the sphere toward the tip.  Therefore, when the tip and the
particle are close enough for the $z$ component of the force to be
positive and to lift the particle off the surface, the lateral force
actually helps bringing the particle in the trap.  Again TE
polarization gives a different result. The lateral force always pushes
the particle away from the tip.  However, since the magnitude of the
(downward) $z$ component of the force is larger than the $x$ component
by a factor of 5, we expect that the sphere is not displaced when the
tip is scanned over it under TE illumination.  Note that apertureless
probes are often used in tapping mode when imaging a surface. This
mode minimizes the lateral motion imparted to the object by the
optical force.

We have shown that a tungsten probe can be used to trap efficiently a
nanometric object above a surface using TM illumination. For
nanomanipulation it is important to assess the stability of the trap
as the probe lifts the particle off the substrate.  Figure~5 shows the
$z$ component of the force when the sphere is fixed at the apex of the
tip and the tip is moved vertically (solid and dashed curves).  The
optical force remains larger than the weight of the particle over a
range of several tens of nanometers.  The particle can therefore be
manipulated vertically as well as horizontally.  Note that the
evolution of the force with the distance to the substrate is linear
rather than exponential. The particle experience a negative gradient
force due the exponential decay of the intensity of the
illumination. At the same time, the particle suffers a positive
gradient force due to the field enhancement at the tip apex, which
also decreases exponentially with $z$ because this enhancement depends
on the intensity of the evanescent illuminating light.  The
competition between these two contributions results in an weak
decrease of the trapping force as the particle is moved away from the
substrate.  If we change the nature, size or shape of the object the
magnitude of the force changes but the conclusions are qualitatively
the same (Fig. 5, curves with symbols).
\begin{figure}[H]  
\begin{center}
\includegraphics*[draft=false,width=80mm]{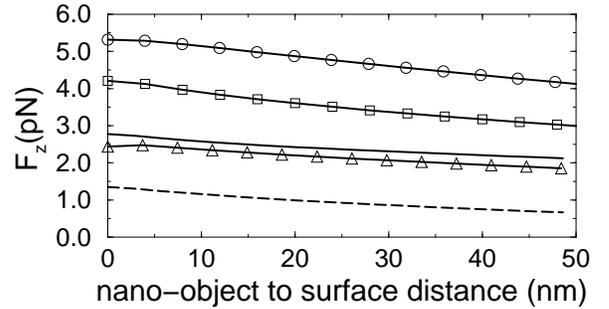}
\end{center}
\caption{$z$ component of the force experienced by a particle as a
function of the distance between the particle and the substrate under
TM illumination. The particle is placed at the apex of the
probe. Solid line : sphere ($\epsilon=2.25$, radius: 10~nm,
$\theta=43^{\circ}$).  Dashed line: sphere ($\epsilon=2.25$, radius:
10~nm, $\theta=50^{\circ}$).  Circles : sphere ($\epsilon=2.25$,
radius: 30~nm, $\theta=50^{\circ}$).  Triangles (force $\times 0.1$):
sphere of gold ($\epsilon=-2.81+3.18i$, radius: 10~nm,
$\theta=50^{\circ}$).  Squares : cube ($\epsilon=2.25$, size: 20~nm,
$\theta=50^{\circ}$).}
\end{figure}

It is fundamental to know whether our procedure also works if several
particles are clustered together. We consider a set of three spheres
(radius 10~nm, permittivity 2.25) aligned along $x$. The probe is
placed above the middle sphere. We account for the multiple scattering
between the three spheres, the substrate and the tip, therefore the
optical binding experienced by the spheres~\cite{chaumet3} is included
in our description. For this configuration, TE illumination again does
not lead to trapping.  For TM illumination, we plot in Fig.~6 the $z$
component of the force experienced by the middle sphere and by those
on the sides as a function of the vertical distance between the probe
and the middle sphere.  For an angle of incidence $\theta=43^{\circ}$,
the $z$ component of the force, although largest for the middle
sphere, is positive for all the spheres. This could be a problem if
one wanted to manipulate only one particle.  The central particle can
be selectively trapped by increasing the angle of incidence of the
illuminating beams to tighten the trap in the $x$ direction (Fig.~4).
In Fig.~6 we see that for $\theta=50^{\circ}$ the optical force
induced by the probe is positive only for the middle sphere. This
remains true for three spheres aligned along $y$.  Figure 7 shows the
extraction of the middle sphere by the tip.  Our calculation shows
that the vertical force experienced by the two side spheres remains
negative when the probe moves away from the substrate.  Therefore, the
spheres on the sides do not hinder the capture of the middle sphere.
\begin{figure}[H]  
\begin{center}
\includegraphics*[draft=false,height=80mm,angle=90]{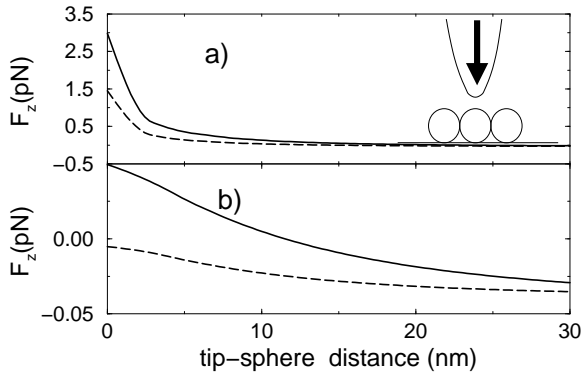}
\end{center}
\caption{Force along $z$ versus vertical position of the probe for
three spheres aligned along $x$. The probe is centered over the middle
sphere.  Solid line $\theta=43^{\circ}$. Dashed line
$\theta=50^{\circ}$.  a) Force on the middle sphere. b) Force
experienced by the spheres on the sides.}
\end{figure}
\begin{figure}[H]  
\begin{center}
\includegraphics*[draft=false,height=80mm,angle=90]{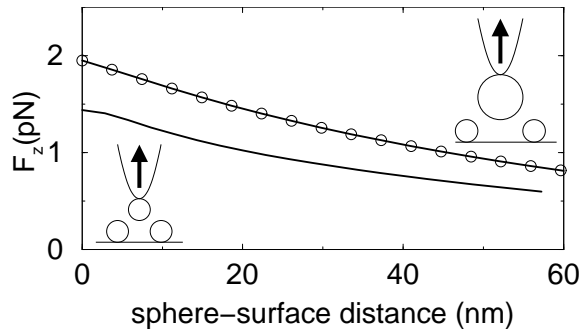}
\end{center}
\caption{Force along $z$ as experienced by the middle sphere (solid
line) versus vertical position of the middle sphere trapped at the
apex probe ($\theta=50^{\circ}$). Solid line : sphere with a 10~nm
radius.  Circles : sphere with a 20~nm radius.}
\end{figure}

In Fig. 7, the curve with circles shows that the selective trapping
works as well when the middle sphere is twice as big as the others.
Once the chosen particle has been trapped, it can be moved above the
substrate.  Our calculations show that the presence of other particles
on the substrate does not destroy the trap, provided that during the
manipulation the trapped sphere is kept at least 5~nm above the
spheres that are on the substrate. We have checked that if the optical
binding causes the side spheres to move laterally toward the middle
sphere, increasing the angle of illumination still creates a negative
(downward) force on the two side spheres, and the selective capture of
the middle sphere is not hampered.

The procedure to trap a small object with a tungsten tip is therefore
the following: first TE illumination is used while the tip scans the
surface in tapping mode or in constant-height mode if the area under
investigation is small enough. Such modes avoid the displacement of
the particle by the tip.  Once an object has been selected, the probe
is placed above the object and the polarization of the illumination is
rotated to TM. The probe is then brought down over the particle and
captures it. The probe can then move the particle above the substrate,
both horizontally and vertically, to a new position where it can be
released by switching back to TE polarization (note that if, for some
reason, one wishes to move the particle over distances larger than the
size of the illumination spot, one could move \emph{the sample} with
piezoelectric device once the sphere is trapped at the apex of the
tip). The lack of trapping under TE illumination is actually an
important advantage during the imaging (selection) and release phases
of the manipulation. Indeed, under TE illumination, when the tip is
above a particle it actually increases the downward optical force,
which contributes to prevent the tip from sweeping the particle away.

In conclusion, we propose a new method to trap and manipulate
nanometric particles in air above a substrate, using an apertureless
tungsten probe.  The probe is used to scatter two counter-propagating
evanescent waves, generating an optical trap at the apex.  We showed
that an object of a few nanometers, can be selectively trapped and
manipulated.  An interesting extension of this work will be a study of
the influence of different illuminations (e.g. focused beam) and a
systematic study of the influence of the nature of the particle and
the tip on the trapping mechanism. For example, the strong spectral
dependence of the electromagnetic response of metal particles (or
resonance excitation of both dielectric and metallic particles) could
lead to a material selective trapping.

P. C. C. and A. R. would like to thank  St\'ephanie Emonin
for many fruitful discussions.


\end{multicols}


\begin{references}
\bibitem{phase} Because typical laser coherence times are small,
(e.g. 200ps for an Argon laser), a simple calculation shows that a
glass sphere with a radius of 10 nm, trapped by an optical force of
about 4pN, will see spatial fluctuations in the 30 pm range, due to
the laser phase fluctuation. Therefore, in real cases the trapped
particule will only feel the time-averaged trapping potential, without
actually being perturbed by the laser fluctuations.
\end{references}
\end{document}